\documentclass{amia}
\usepackage{lipsum} 
\usepackage{bm}
\usepackage{amsmath} 
\usepackage{graphicx}
\usepackage{subcaption}
\usepackage{natbib} 
\usepackage[dvipsnames]{xcolor}
\usepackage[colorlinks]{hyperref}
\usepackage{subcaption}
\hypersetup{
    linkcolor=blue,
    filecolor=magenta,
    urlcolor=teal,
    citecolor=magenta
    }
\makeatletter
\def\blfootnote{\xdef\@thefnmark{}\@footnotetext}
\makeatother

\begin{document}

\title{Estimation-Theoretic Bias Reduction for Oscillometric Blood Pressure Readings}

\author{Masoud Nateghi$^1$, Reza Sameni$^{1,2}$ }

\institutes{
    $^1$ Department of Biomedical Informatics, Emory University, Atlanta, GA\\ 
    $^2$ Department of Biomedical Engineering, Emory University and Georgia Institute of Technology, Atlanta, GA
}

\maketitle

\blfootnote{Manuscript accepted for presentation at the AMIA 2026 Annual Symposium,
Dallas, Texas, USA, November 7--11, 2026.}


\section*{Abstract}

\textit{Oscillometry is the standard method for non-invasive, cuff-based blood pressure (BP) measurement, but it introduces systematic errors that may impact clinical accuracy. This study investigates the sources of these errors—primarily the limitations of oscillometry itself and respiration-induced fluctuations—using BP waveform data from the MIMIC database. Oscillometry tends to underestimate systolic BP and overestimate diastolic BP, while respiration introduces cyclical variations that further degrade measurement precision. To mitigate these effects, we propose an estimation-theoretic framework employing least squares (LS) and maximum likelihood (ML) methods for correcting both single and multiple BP measurements. LS estimation corresponds to conventional multi-measurement averaging protocols, whereas the ML approach incorporates prior knowledge of measurement errors, offering improved performance. Our results demonstrate that averaging repeated BP measurements and incorporating prior knowledge of measurement error statistics can improve the accuracy of non-invasive BP estimation.}

\section*{Introduction}
\label{sec:introduction}
Cardiovascular disease (CVD) is the leading global cause of death, accounting for 19.8 million deaths annually~\cite{WHO_CVDs_2025}. Blood pressure (BP) measurement is essential for CVD screening and early intervention~\cite{Rastegar2019, Muntner2019}. Despite significant technological advances in BP measurement devices and regulatory validation, BP readings remain prone to systematic errors—arising from heuristic, device-specific algorithms, variability in patient physiology, and procedural flaws such as improper preparation, cuff misplacement, or insufficient readings—that can impact accuracy in real-world conditions~\cite{Mousavi2024, Palatini2025}. Even validated and well-calibrated devices may yield inaccurate readings if proper measurement procedures are not followed~\cite{Elias2021, Muntner2019, Shimbo2020}. These inaccuracies can significantly affect reliability: a U.S. study estimated that a 5\,mmHg error could misclassify BP in 48 million individuals annually~\cite{Jones20031027, Argha2022}. Overestimation may lead to overtreatment and higher costs~\cite{Jones20031027, Kumar2023}, while underestimation risks missing hypertension diagnoses~\cite{Psaty1997, Kumar2023}. Despite widespread use, non-invasive methods like auscultation and oscillometry often diverge from intra-arterial measurements, the clinical gold standard~\cite{Mousavi2024, Finnie1984, Duncombe2017}.

Addressing BP measurement errors requires a multifaceted approach that considers human factors at the point of care and advances in hardware and software. A promising strategy is post-measurement calibration (either by the device firmware or in software), where known biases are corrected to better approximate true BP. For example, machine learning models trained on retrospective datasets combining cuff-based and invasive readings could learn to adjust for systematic errors. However, this approach is limited by scarce simultaneous invasive and non-invasive data, and by variability across subjects and sessions. These challenges underscore the need for a robust, theoretically grounded framework to ensure reliable bias correction. In this work, we develop and evaluate such a framework.

We introduce an estimation-theoretic framework designed to improve the accuracy of oscillometric blood pressure measurements after data acquisition. These algorithms are suitable for integration into the software of BP monitoring devices, supporting both discrete readings and continuous measurements from the arm or fingers. By analyzing intra-arterial BP recordings from the MIMIC database~\cite{MIMIC}, we identified two primary sources of error: respiratory-induced short-term fluctuations and fundamental limitations of the oscillometric method—such as rapid cuff deflation and imprecise timing—that typically cause underestimation of systolic and overestimation of diastolic pressures~\cite{Mousavi2024, Bui2023, Picone2017}. To systematically address these biases, we developed a model of the oscillometric measurement process vs intra-arterial BP data, considered ``ground truth'', enabling simulation and detailed analysis of key error mechanisms. We then assessed the performance of various estimation techniques, including least squares (LS) and maximum likelihood (ML), in compensating for these errors. To implement these frameworks, we conducted a population study using the MIMIC dataset to obtain the necessary hyper-parameters for each framework.

The proposed framework provides valuable insights into the underlying causes of BP measurement biases and highlights how the averaging of multiple readings can effectively reduce these errors. We demonstrate that increasing the number of measurements significantly reduces the variance in BP estimates across all methods, thereby enhancing the accuracy and confidence of the results (reducing variances).

\section*{Oscillometric BP measurement}
\label{sec:model}
BP is a critical measure of cardiovascular health. When recorded continuously, BP has a waveform with recurring peaks and troughs aligned with each heartbeat, representing pressure changes throughout the cardiac cycle~\cite{guyton2021}. Discrete BP measurements summarize this waveform using key indicators: systolic blood pressure (SBP), diastolic blood pressure (DBP), pulse pressure (PP), and mean arterial pressure (MAP). SBP corresponds to the highest pressure during ventricular contraction; DBP is the lowest pressure over a cycle; PP is the difference between SBP and DBP; and MAP=(2DBP+SBP)/3 is a weighted mean between SBP and DBP~\cite{guyton2021}. 

Oscillometry remains the leading technique for non-invasive BP measurement. It uses a cuff, usually wrapped around the upper arm over the brachial artery, to measure BP indirectly. The cuff is first inflated to a pressure higher than the expected systolic pressure (e.g., 180\,mmHg) and then slowly deflated below the diastolic pressure (50\,mmHg or below)~\cite{Drzewiecki1994, Liu2012}. At sufficiently high cuff pressure, the artery is occluded and blood flow stops. As the cuff pressure drops below the systolic level, small pulses from the heartbeat cause vibrations in the artery. These vibrations increase as cuff pressure decreases, reaching their maximum near MAP, and then fade as pressure falls below the diastolic level, allowing blood to flow freely. These arterial vibrations can be detected in two ways. In the traditional auscultatory method, they manifest as Korotkoff sounds that are detected using a stethoscope to determine systolic and diastolic pressures. In contrast, oscillometric devices capture the resulting cuff pressure oscillations using pressure sensors and analyze their envelope during cuff deflation to estimate MAP and subsequently derive systolic and diastolic pressures using empirical algorithms~\cite{Liu2012, Sharman2022}.

Current BP measurement algorithms are mostly heuristic and prone to systematic errors. Most commercially available oscillometric devices rely on proprietary algorithms that are not publicly disclosed, making it difficult to systematically evaluate or reproduce their performance~\cite{Palatini2025, Kumar2023}. Studies show they often underestimate SBP and overestimate DBP~\cite{Mousavi2024, Bui2023, Picone2017}. These biases can have serious clinical consequences, especially for vulnerable populations like pregnant women, older adults, or cardiovascular patients.

To quantify oscillometric biases, we used continuous arterial pressure data from the MIMIC dataset and simulated the interaction with an external cuff pressure that decreases at a typical clinical deflation rate of 2–3\,mmHg/s. By finding the points where the cuff pressure intersects the arterial waveform, we identified simulated SBP and DBP achieved using oscillometry. These measurements were further used in our error analysis and bias correction framework evaluations.

\section*{Methods}
\subsection*{A data model for oscillometric BP measurement}
The relationship between the true BP, respiration effects, and oscillometric measurement errors can be modeled using a linear additive framework, identifying respiration and inherent oscillometric errors as the two main sources of bias in BP readings. This model can be formulated as:
\begin{equation}
    \mathbf{x}_k = \bm{\theta} + \mathbf{n}_k
    \label{eq:model1}
\end{equation}
where $\mathbf{x}_k=[s_k, d_k]^T$ represents the $k$-th measurement of SBP and DBP; $\bm{\theta}=[\text{SBP}, \text{DBP}]^T$ is the true, unbiased BP vector; $\mathbf{n}_k$ represents measurement error (combining repiratory effect and inherent oscillometric error), and $k$ is the measurement index. In clinical settings, multiple measurements may be taken to accurately estimate BP. For the current study, we assume the `true' BP ($\theta$) remains constant over these measurements and that measurement noise is independent, identically distributed (i.i.d.), and does not depend on the observed data.

Next, we explore two approaches to estimate the true BP $\theta$. First, a least squares-based method that makes no assumptions about the distributions of $\theta$ or the noise; second, a maximum likelihood-based method that assumes a known distribution for the noise. We will explain how each approach can be applied in various BP measurement contexts.
\subsection*{Estimation frameworks for BP measurement bias correction}
\label{sec:formula}
\paragraph{Least Squares Error (LS):}
LS and its variants, such as weighted LS, are widely used for parameter estimation when little or no prior knowledge exists about the error or parameter distributions. LS relies only on the data model and aims to minimize the sum of squared differences between the observed and predicted values. In our case, applying LS to the data model \eqref{eq:model1} leads to the following estimate~\cite{Kay93}:

\begin{equation}
    \hat{\bm{\theta}}_{\text{LS}}=\frac{1}{N}\sum_{k=1}^{N} \mathbf{x}_k = \bar{\mathbf{x}}
    \label{eq:LS}
\end{equation}
\paragraph{Maximum Likelihood (ML):}
ML estimation determines parameter values by maximizing $f(\mathbf{x}|\bm{\theta})$, yielding the most probable $\bm{\theta}$ given the observed data $\mathbf{x}$. While ML does not require specifying the distribution of the parameters $\bm{\theta}$, it does require knowledge of the observation noise distribution. Herein, we assume a Gaussian-distributed noise vector $\mathbf{n}_k \sim \mathcal{N}(\bm{\mu}_n, \mathbf{C}_n)$ that is independent and identically distributed across $N$ observations. The parameters of this Gaussian distribution (which aggregates all sources of bias and error) can be estimated from population studies and device specifications, comparing accurate intra-arterial BP measurements with cuff-based measurements. This Gaussian assumption serves as our working assumption to facilitate ML model derivation. More generally, it requires validation using real measurement data. Under this assumption, the ML estimator for true BP values is derived as~\cite{Kay93}:
\begin{equation}
    \hat{\bm{\theta}}_{\text{ML}}=\frac{1}{N}\sum_{k=1}^{N} \mathbf{x}_k-\bm{\mu}_n=\bar{\mathbf{x}}-\bm{\mu}_n
    \label{eq:MLE}
\end{equation}

\subsection*{Model Parameter Estimation}
\label{subsec:model_param_est}

Among the BP estimation methods considered, LS approach does not require any prior statistical assumptions and can be directly applied to the measurement data. In contrast, ML method requires knowledge of the distribution of measurement noise. 
In the ML framework, to estimate the mean \(\bm{\mu}_n\) and covariance \(\mathbf{C}_n\) of the measurement noise, we calculate the difference between the simulated oscillometric BP values and the ground truth intra-arterial measurements from the MIMIC dataset. These differences represent the combined measurement noise and are used to estimate the statistical parameters required for ML estimation.


\subsection*{The Proposed BP Estimation Framework in Practice}
\label{subsec:practical_implementation}

The LS and ML estimation frameworks differ in terms of their assumptions and intended applications. The LS method does not rely on any statistical assumptions about the measurement noise or the true BP values. It uses only the observed data and can be applied directly without requiring additional parameters. In practice, the LS estimator reduces to the arithmetic mean of multiple BP measurements. Therefore, commonly used clinical protocols that estimate BP by averaging repeated readings (e.g., taking the mean of three measurements) are mathematically equivalent to applying the LS framework.

In contrast, the ML framework incorporates statistical information about the measurement error. It assumes a known distribution for the noise and uses the mean error \(\bm{\mu}_n\) to adjust the observed measurements. This makes ML particularly useful in settings where the characteristics of measurement errors—such as device-specific oscillometric biases—are known or can be estimated. As a result, ML is well-suited for device calibration or improving accuracy in clinical environments where measurement error distributions are well characterized.

\section*{Dataset}
\label{sec:dataset}

To evaluate the proposed BP estimation and correction methods, we use the publicly available MIMIC database~\cite{MIMIC}. This dataset contains deidentified health records from over 90 patients admitted to intensive care units (ICUs). It combines data from bedside monitors and clinical documentation, offering a comprehensive, multimodal view of each patient. The physiological recordings typically last at least 20 hours, with many extending beyond 40 hours.
The dataset includes several vital signals, such as electrocardiogram (ECG), arterial blood pressure (ABP), pulmonary arterial pressure (PAP), central venous pressure (CVP), and fingertip plethysmograph (PLE). Among 72 subjects, 68 had available intra-arterial BP waveforms, which were used for this study.

\section*{Preprocessing}
\label{sec:preprocessing}

The continuous BP signals were divided into 60\,s segments, and a set of threshold-based rules was used to filter out defective segments. Segments with BP values below 25\,mmHg or above 180\,mmHg were removed, except for hypertensive subjects with elevated BP values, where an upper limit of 230\,mmHg was used. These thresholds help eliminate segments likely affected by measurement drift or artifacts caused by blood clutter (e.g., clots or air bubbles) in the arterial line, which can distort pressure readings and typically require flushing to correct~\cite{Saugel2020}. We also automatically excluded saturated segments, where the signal remained constant for 500\,ms or longer. After cleaning, the global maxima and minima within each valid segment were used as the ground truth SBP and DBP values, respectively.

\subsection*{Results}
\subsubsection*{Respiration: a significant source of bias in BP monitoring}
\begin{figure}[tb]
    \centering

    \begin{subfigure}[t]{0.49\textwidth}
        \centering
        \includegraphics[
            trim=0cm 0cm 0cm 0cm,
            clip=true,
            width=0.9\linewidth
        ]{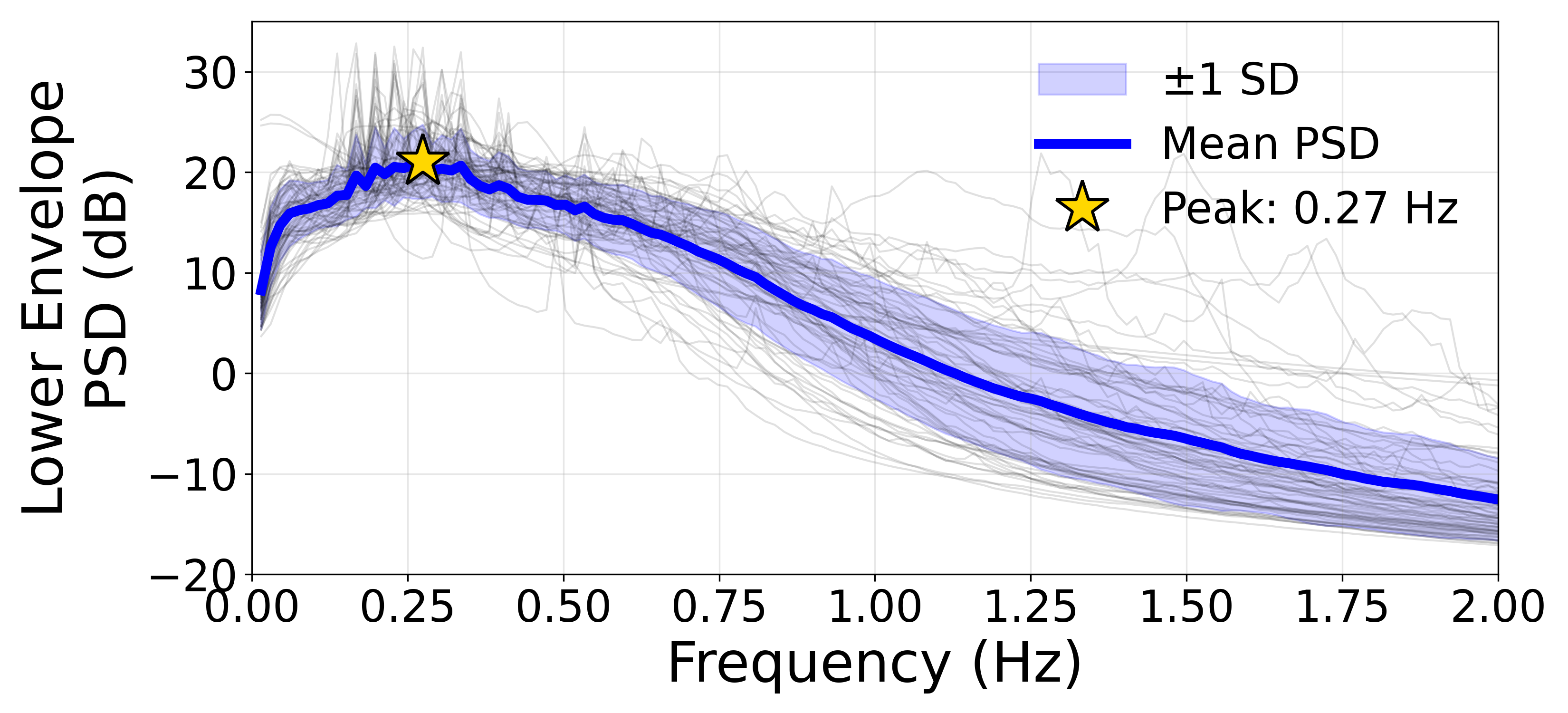}
        \caption{}
        \label{fig:PSD_lower}
    \end{subfigure}
    \hfill
    \begin{subfigure}[t]{0.49\textwidth}
        \centering
        \includegraphics[
            trim=0cm 0cm 0cm 0cm,
            clip=true,
            width=0.9\linewidth
        ]{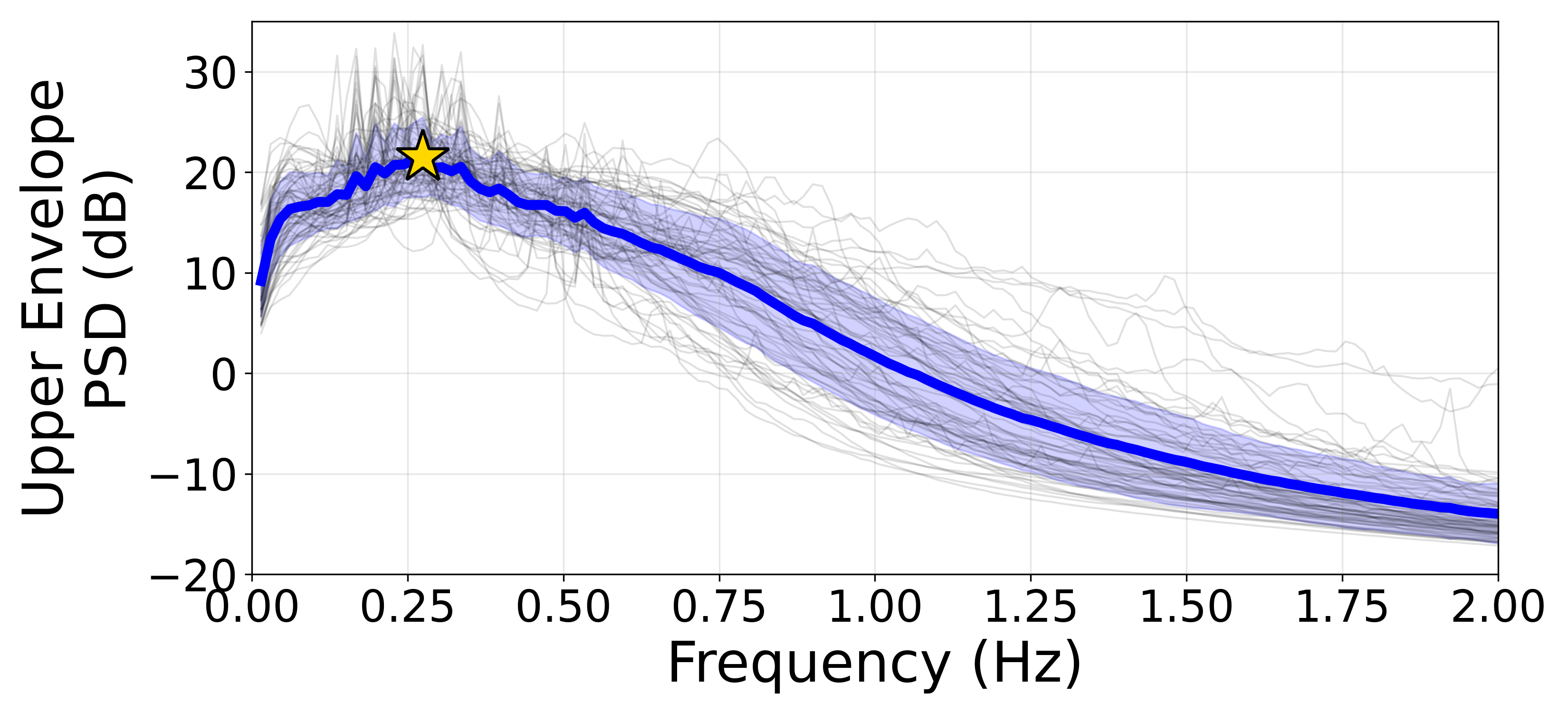}
        \caption{}
        \label{fig:PSD_upper}
    \end{subfigure}

    \caption{PSD of the (a) lower and (b) upper BP waveform envelopes from individual MIMIC recordings (light gray), together with the mean PSD across subjects (blue) and the corresponding $\pm1$ standard deviation region (shaded). The dominant spectral peak (asterisk) occurs near 0.27\,Hz, corresponding to the typical respiration frequency range, supporting respiration as a major contributor to short-term BP variability.}

    \label{fig:PSD}
\end{figure}

Respiration is known to introduce slow fluctuations in the arterial blood pressure (BP) waveform, which can influence oscillometric BP measurements. Since the MIMIC dataset does not contain independent respiratory recordings for most subjects, we instead investigated respiratory influence indirectly by analyzing periodic variations in the BP waveform envelopes.
To extract the upper and lower envelopes of the BP waveform, we applied a 10\,s moving average filter to estimate the instantaneous MAP, assuming that MAP varies smoothly over this time scale. This smoothed signal served as a dynamic threshold that suppressed local minima and enhanced peak prominence. By setting values below this threshold to zero and applying peak detection, we identified SBP values for each heartbeat and used them to interpolate the upper envelope. The exact process was repeated on the inverted BP signal to extract the lower envelope.
Using these envelopes, we computed the power spectral density (PSD) of each subject's envelope signal and averaged them across all subjects to identify dominant frequencies. As shown in Figure~\ref{fig:PSD}, the average PSD of both the upper and lower BP envelopes exhibits a dominant spectral peak near 0.27\,Hz, within the typical respiration frequency range of approximately 0.2--0.4\,Hz. This observation suggests that respiration consistently modulates the BP waveform across subjects and represents a major source of short-term oscillometric BP variability.



\subsubsection*{Inherent limitation of oscillometry}
\label{sec:oscillometry_limits}
To simulate oscillometric BP measurement, we started with an initial cuff pressure of 180\,mmHg (or 230\,mmHg for subjects with elevated BP) that decreased linearly at a rate of 2.5\,mmHg/s (3.5\,mmHg/s for elevated BP subjects). This ensured the cuff pressure was always above the SBP values, as required by the oscillometric method. We then found the intersection points between this linearly decreasing cuff pressure and the continuous BP signal for each 1-minute segment. The first intersection was taken as the estimated SBP and the last as the estimated DBP, representing noisy oscillometric measurements.

Across the MIMIC dataset, these oscillometric estimates systematically underestimated SBP by an average of 9.29\allowbreak\,mmHg and overestimated DBP by 6.29\,mmHg. These biases informed the selection of the noise mean $\bm{\mu}_n$ used in our estimation models. The covariance matrix $\mathbf{C}_n$ had non-zero off-diagonal elements, showing correlation between SBP and DBP errors, a phenomenon previously reported in large population studies~\cite{Mousavi2024b}.

\begin{figure}[tb]
    \centering
    \begin{subfigure}{0.48\linewidth}
        \centering
        \includegraphics[
            trim=0cm 0cm 0cm 0cm,
            clip=true,
            width=\linewidth
        ]{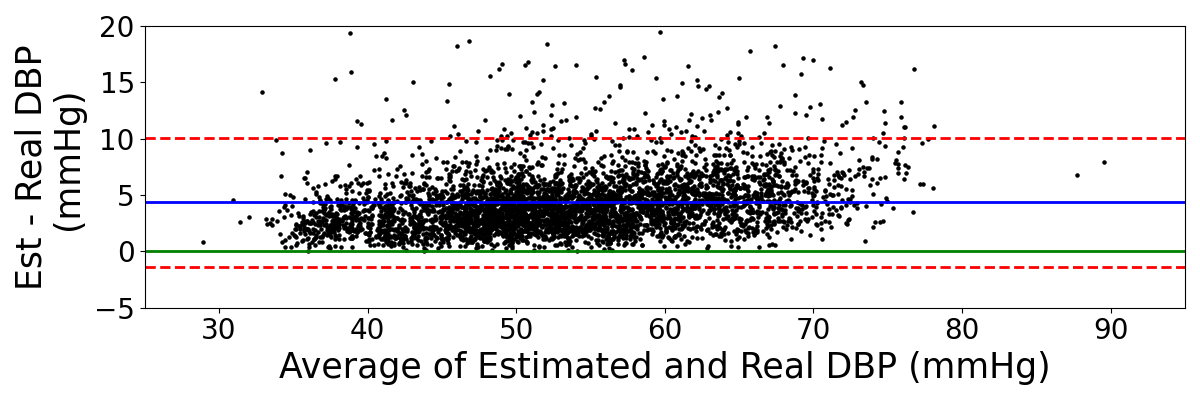}
        \caption{DBP estimation error, deflation rate: $2.5$\,mmHg/s}
        \label{fig:bland_a}
    \end{subfigure}
    \hfill
    \begin{subfigure}{0.48\linewidth}
        \centering
        \includegraphics[
            trim=0cm 0cm 0cm 0cm,
            clip=true,
            width=\linewidth
        ]{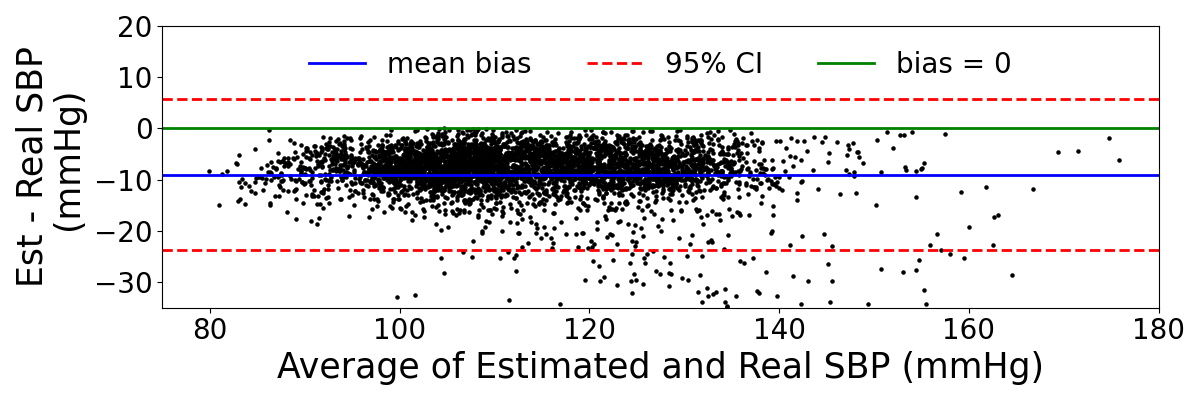}
        \caption{SBP estimation error, deflation rate: $2.5$\,mmHg/s}
        \label{fig:bland_b}
    \end{subfigure}

    \vspace{2em}

    \begin{subfigure}{0.48\linewidth}
        \centering
        \includegraphics[
            trim=0cm 0cm 0cm 0cm,
            clip=true,
            width=\linewidth
        ]{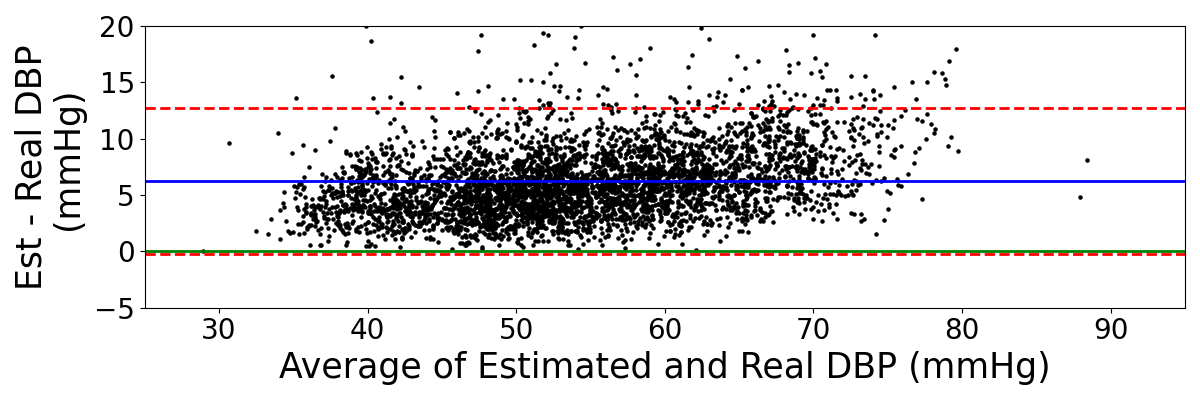}
        \caption{DBP estimation error, deflation rate: $10$\,mmHg/s}
        \label{fig:bland_c}
    \end{subfigure}
    \hfill
    \begin{subfigure}{0.48\linewidth}
        \centering
        \includegraphics[
            trim=0cm 0cm 0cm 0cm,
            clip=true,
            width=\linewidth
        ]{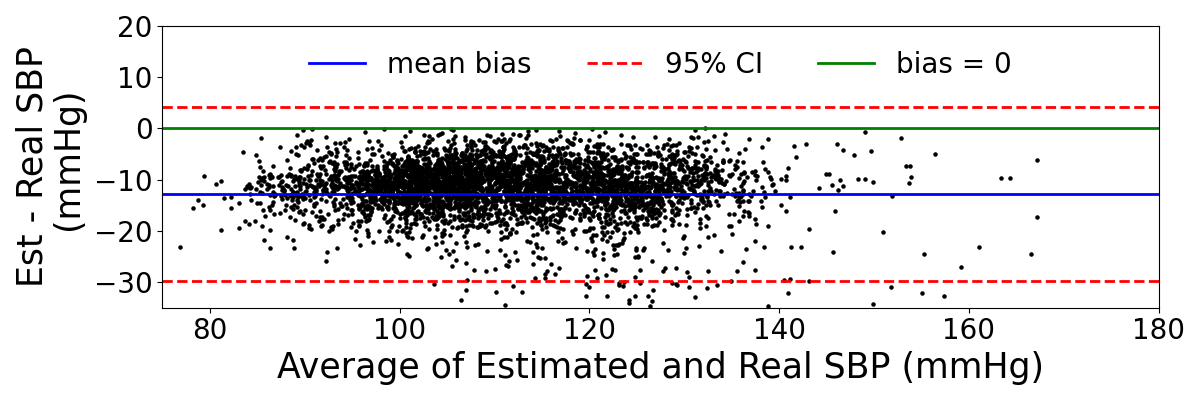}
        \caption{SBP estimation error, deflation rate: $10$\,mmHg/s}
        \label{fig:bland_d}
    \end{subfigure}
    \caption{Bland--Altman analysis of simulated oscillometric SBP and DBP measurements for subject 039 under two cuff deflation rates. Both cases exhibit underestimation of SBP and overestimation of DBP. Faster deflation increases both the bias and variability of the estimated BP values.}

    \label{fig:bland_altman}
\end{figure}

Figures~\ref{fig:bland_a} and~\ref{fig:bland_b} present Bland--Altman analyses comparing simulated oscillometric BP measurements against the ground-truth intra-arterial values for a typical cuff deflation rate of 2.5\,mmHg/s. In particular, the mean differences in the plots are shifted away from zero, with SBP differences appearing predominantly negative (indicating systematic underestimation) and DBP differences appearing predominantly positive (indicating systematic overestimation). Our simulation suggests that a major cause is the cuff pressure's inability to align exactly with the true SBP and DBP points on the arterial waveform. The cuff deflation rate is critical here: faster deflation reduces the chance of accurately capturing the true peaks and troughs, increasing measurement error, as shown in Figures~\ref{fig:bland_c} and~\ref{fig:bland_d}. Slower cuff deflation decreases this bias, but even with very slow deflation, respiratory effects remain a significant source of error.



\subsubsection*{Bias-correction frameworks}
\label{sec:performance_comparison}
To mitigate the biases described in the previous sections, the methods given in~\eqref{eq:LS} and~\eqref{eq:MLE} can be applied to either a single measurement or multiple BP measurements. Then, we evaluated their performance by comparing the corrected BP values from each framework to the ground-truth BP values.

\begin{figure}[tb]
    \centering
    \hfill
    \begin{subfigure}[t]{0.49\textwidth}
        \centering
        \includegraphics[
            trim=0cm 0cm 0cm 0cm,
            clip=true,
            width=0.85\linewidth
        ]{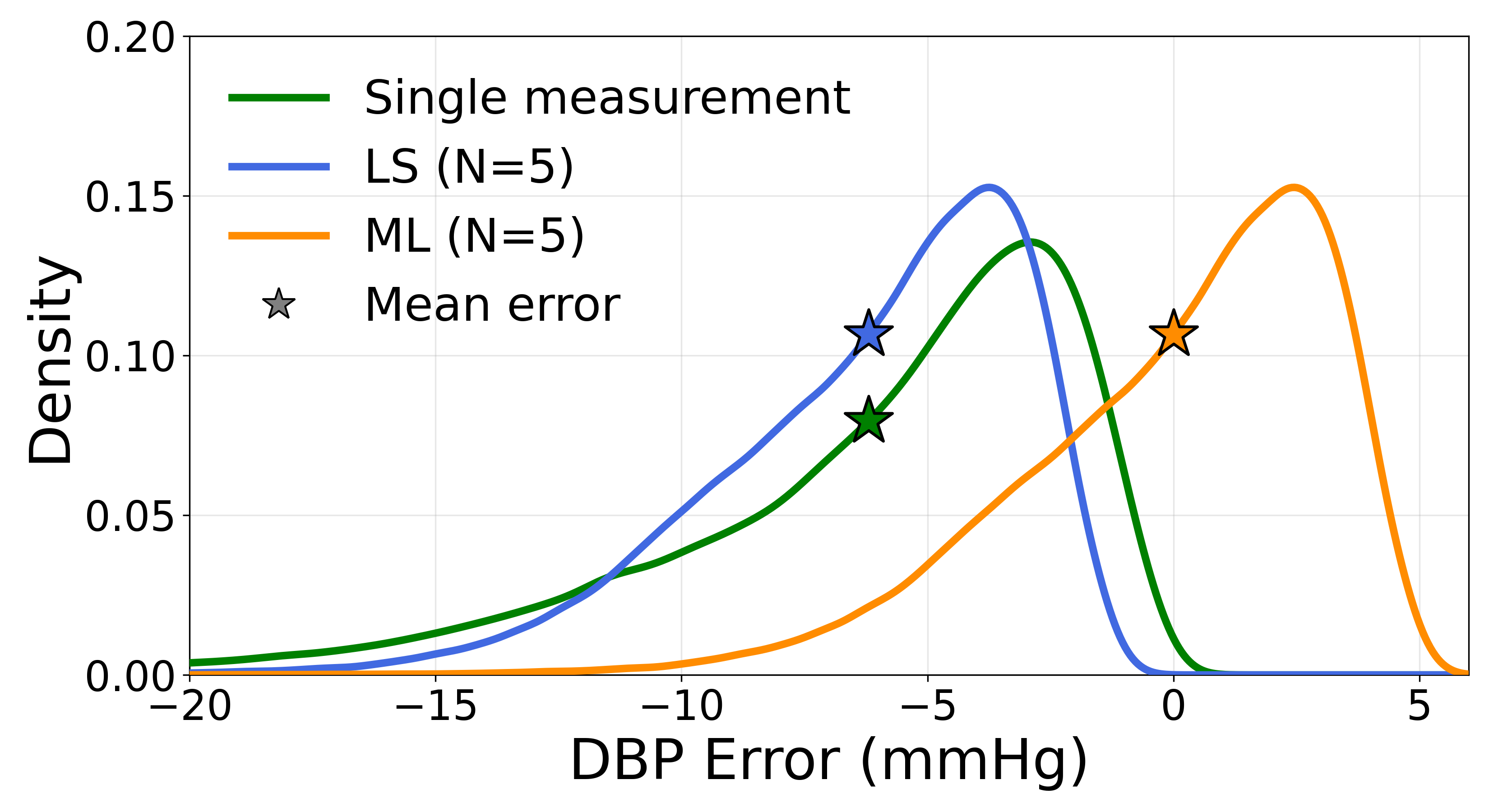}
        \caption{DBP estimation error distributions.}
        \label{fig:pdf_sbp}
    \end{subfigure}
    \hfill
    \begin{subfigure}[t]{0.49\textwidth}
        \centering
        \includegraphics[
            trim=0cm 0cm 0cm 0cm,
            clip=true,
            width=0.85\linewidth
        ]{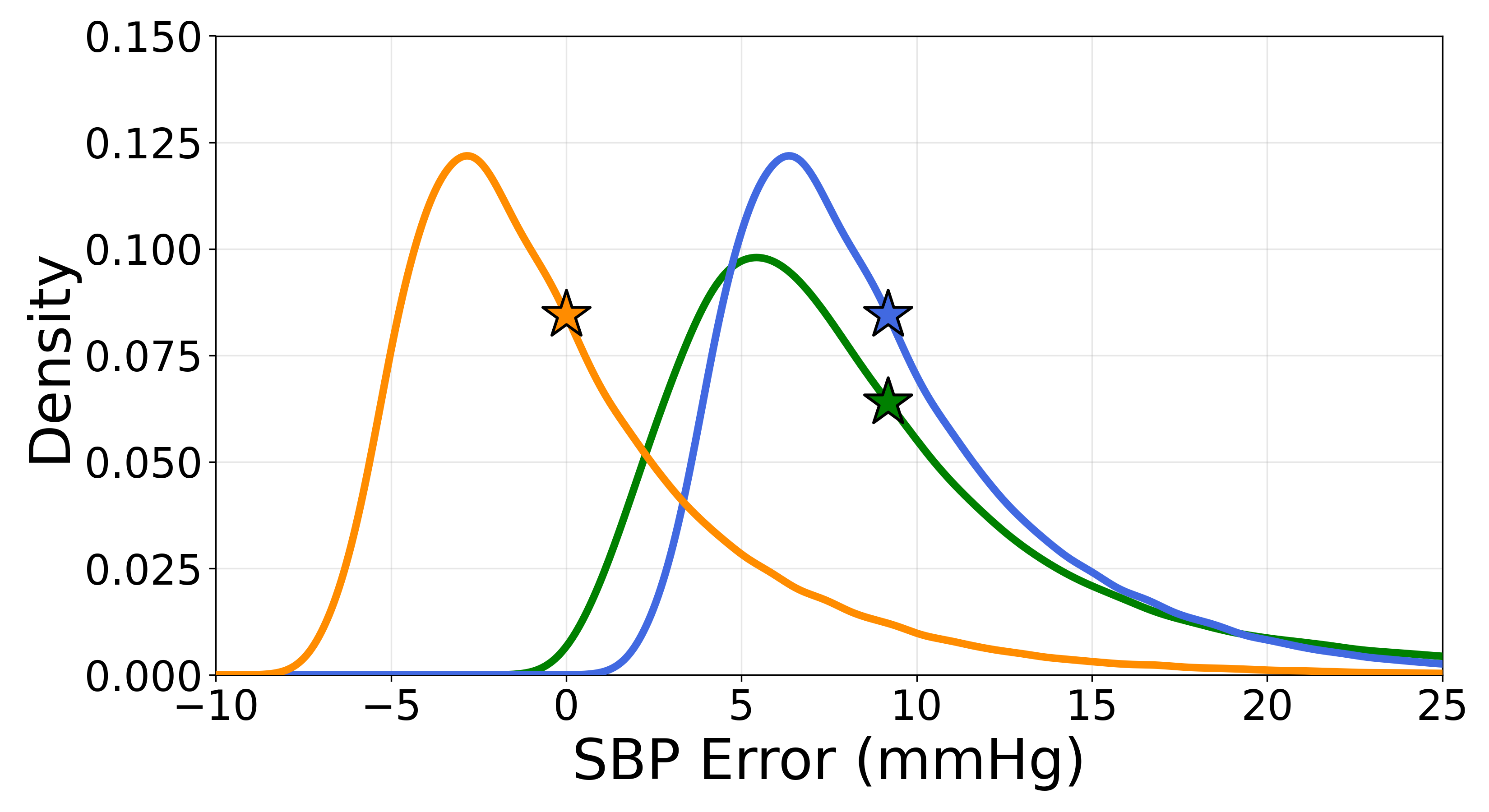}
        \caption{SBP estimation error distributions.}
        \label{fig:pdf_dbp}
    \end{subfigure}
    \hfill
    \caption{Normalized histograms and kernel density estimates of SBP (a) and DBP (b) estimation errors across all subjects in the MIMIC dataset using three approaches: a single measurement, LS with $N=5$ measurements, and ML with $N=5$ measurements. Asterisks denote the mean estimation error of each method. Multi-measurement approaches substantially reduce error variability compared to single measurements. However, LS exhibits systematic bias, reflected by shifted error distributions, whereas ML partially compensates for this bias by incorporating measurement error statistics, resulting in distributions that are more centered around zero.}

    \label{fig:pdf_err}
\end{figure}

Figure~\ref{fig:pdf_err} shows the probability density functions (PDFs) of SBP and DBP estimation errors after bias correction using multiple measurements, compared to a single measurement without any correction, across all subjects of the MIMIC dataset. The LS framework tends to underestimate SBP and overestimate DBP. This is expected, since LS averages the biased measurements directly without any correction, and is reflected in the non-zero means of the LS error PDFs for both SBP and DBP. In contrast, the ML framework improves estimation performance by incorporating the correction term $\bm{\mu}_{n}$, which accounts for the systematic measurement bias and sets the mean of the ditribution to zero. 

Table~\ref{tab:mae_std_results} summarizes the mean absolute error (MAE) and standard deviation (STD) for these frameworks using both single and multiple measurements. Averaging multiple measurements significantly reduces variance in BP estimates. Under the assumption of no model mismatch and i.i.d.\ measurements, the covariance of the error should decay proportionally to $1/N$~\cite{Kay93}. 


\begin{table}[H]
\centering
\caption{Mean absolute error (MAE) and standard deviation of error for bias-correction frameworks across all MIMIC dataset subjects, comparing $N=1$ and $N=5$ measurements.}
\label{tab:mae_std_results}
\begin{tabular}{lcccc}
\hline
\textbf{Method} & $\mathbf{N}$ & \textbf{SBP MAE} & \textbf{DBP MAE} \\
\hline
LS & $1$ & $9.29 \pm 7.39$ & $6.29 \pm 4.70$ \\
LS & $5$ & $9.29 \pm 4.92$ & $6.29 \pm 3.26$ \\
\hline
ML & $1$ & $4.96 \pm 7.39$ & $3.51 \pm 4.70$ \\
ML & $5$ & $4.96 \pm 4.92$ & $3.51 \pm 3.26$ \\
\hline
\end{tabular}
\end{table}

\section*{Discussion}
Non-invasive BP measurement methods, particularly oscillometry, are known to introduce biases when compared to invasive techniques. Our analysis demonstrated that this bias is intrinsic to the oscillometric measurement process itself. During cuff deflation, oscillometric devices estimate systolic and diastolic pressures indirectly by identifying the intersections between the arterial pressure waveform and a continuously decreasing external cuff pressure. Because the cuff pressure decreases over time, the detected intersection points generally occur before the true systolic peak and after the true diastolic trough of the arterial waveform. As a result, oscillometric measurements tend to underestimate SBP and overestimate DBP. Furthermore, faster cuff deflation rates reduce the likelihood of accurately capturing the true systolic and diastolic extrema, thereby increasing both measurement bias and estimation variance.

In addition to this systematic bias, our analysis showed that respiration introduces slow periodic fluctuations in the arterial BP waveform. These fluctuations modulate both the systolic and diastolic envelopes of the signal and therefore contribute additional variability to oscillometric BP measurements. As the arterial waveform varies with respiration during cuff deflation, the resulting intersection points between cuff pressure and the arterial waveform also vary, leading to increased variability in the estimated systolic and diastolic pressures. The PSD analysis consistently identified dominant frequency components within the typical respiratory frequency range across subjects, supporting respiration as a major physiological source of short-term oscillometric BP variability.

To mitigate these biases, two estimation frameworks were proposed and evaluated. The LS approach aligns closely with standard clinical practices, which typically involve reporting a single measurement or the average of multiple readings. Figure~\ref{fig:pdf_sbp} shows that, for single measurements (LS with $N=1$), the SBP error distribution is shifted toward positive values, whereas in Figure~\ref{fig:pdf_dbp} the DBP error distribution is shifted toward negative values, indicating systematic underestimation of SBP and overestimation of DBP. The LS framework for $N>1$ retains this systematic bias because it directly averages biased measurements without explicitly compensating for the underlying error distribution. Consequently, LS-based estimates continue to underestimate SBP and overestimate DBP. Nevertheless, increasing the number of measurements reduces estimation variance, as reflected by the narrower error distributions observed for $N=5$ compared to single measurements. Under the assumptions of the proposed model, this variance reduction is expected to scale approximately proportionally to $1/N$.

In contrast, the ML framework incorporates prior knowledge of the measurement error distribution to compensate for systematic oscillometric bias. As shown in Figure~\ref{fig:pdf_err}, the ML error distributions are centered closer to zero for both SBP and DBP, indicating substantial reduction of systematic bias. Similar to LS, increasing the number of measurements also reduces the variance of the ML estimates. These findings suggest that repeated BP measurements primarily improve precision by reducing variance, whereas incorporation of prior statistical knowledge regarding oscillometric error characteristics is required to mitigate systematic bias and improve overall estimation accuracy.


An important observation is that oscillometric errors affect SBP and DBP in opposite directions. Since MAP is computed as a weighted combination of these quantities $(\text{MAP} = \frac{2}{3}\text{DBP} + \frac{1}{3}\text{SBP})$, these opposing errors partially cancel when estimating MAP. As a result, MAP estimates tend to be less sensitive to oscillometric bias than SBP or DBP individually. This observation suggests that MAP may serve as a more robust BP biomarker when using oscillometric measurements, particularly in settings where measurement variability and oscillometric bias are difficult to avoid.

From a clinical and informatics perspective, the proposed framework demonstrates how statistical modeling and repeated measurements can improve the reliability of routine non-invasive BP monitoring without requiring changes to existing hardware. Because oscillometric BP devices are widely used in outpatient clinics, hospitals, ICUs, and home monitoring systems, even modest systematic errors can influence hypertension diagnosis, treatment decisions, and longitudinal patient monitoring. The proposed estimation frameworks therefore provide a principled basis for future software-based calibration, bias-correction, and decision-support approaches aimed at improving the accuracy and reliability of non-invasive BP assessment.

\subsection*{Limitations}
We acknowledge some limitations in the current framework, which should be addressed in future research. 
This study used simulated oscillometric BP values derived from intra-arterial waveforms rather than actual simultaneous cuff-based measurements. This was because datasets containing both intra-arterial and simultaneous oscillometric recordings are extremely scarce and difficult to acquire in practice. Simulating BP from ground truth intra-arterial data enables controlled analysis of key oscillometry error mechanisms, particularly those related to oscillometric bias and respiratory fluctuations. However, we acknowledge that simulated measurements may not fully reflect real-world complexities such as device-specific (proprietary) software calibrations, cuff placement variability, patient movement, or environmental noise. Future studies should therefore evaluate the proposed framework using simultaneous cuff-based and invasive arterial-line measurements collected in real-world clinical settings.

The proposed estimation framework relies on a linear additive model with independent, identically distributed Gaussian noise. These assumptions support analytical tractability and closed-form estimators but may not fully represent the non-Gaussian or temporally correlated nature of physiological signals. Moreover, the model assumes that true BP remains constant across multiple measurements, which may not always be valid in clinical settings. While this simplifies the estimation process, future extensions should incorporate extended Bayesian frameworks to allow modeling BP as random variables, or incorporate dynamic models, such as Kalman filtering, to account for temporal BP variability and non-stationary measurement dynamics.

The dataset used in this study did not provide detailed demographic and clinical information. Therefore, the current framework did not incorporate patient-specific factors such as age, BMI, body composition, or severity of illness, since the primary focus was on correcting systematic oscillometric measurement errors. These physiological and clinical factors are known to influence arterial compliance, vascular dynamics, and oscillometric BP measurement characteristics, and may therefore affect both oscillometric bias and the performance of the proposed estimation frameworks across different patient subgroups. In real-world applications, such variability could potentially be incorporated by estimating individualized or subgroup-specific measurement error distributions within the proposed estimation framework. Incorporating such variability is a potential direction for future personalized extensions, particularly within a Bayesian framework.

Further research should also explore broader validation across diverse patient populations and clinical environments, as well as integration with device-level BP bias correction algorithms. Additionally, the joint behavior of SBP and DBP, including both physiological correlations and measurement errors, deserves further investigation. Finally, extending this estimation framework to focus on secondary metrics such as pulse pressure and mean arterial pressure may provide further insights for enhancing cardiovascular monitoring and diagnostic accuracy.

\section*{Conclusion}
This study examined systematic biases in noninvasive BP measurements, emphasizing limitations of oscillometric methods and respiratory influences. These factors typically lead to an underestimation of SBP and an overestimation of DBP.
To address these issues, we evaluated two estimation strategies: least squares and maximum likelihood. While least squares aligns with standard clinical practice, it does not correct for bias. The maximum likelihood method, which incorporates prior error statistics, showed superior performance in reducing systematic biases.
Averaging multiple measurements further improved accuracy and reduced estimation variance across all methods. DBP measurements were generally more accurate than SBP, and MAP offers more robust alternative to individual SBP or DBP values. Overall, incorporating measurement error information and leveraging repeated observations substantially enhanced the accuracy of noninvasive BP estimation.

\makeatletter
\renewcommand{\@biblabel}[1]{\hfill #1.}
\makeatother

\bibliographystyle{vancouver}
\bibliography{References}  

\end{document}